\documentclass[preprint,showpacs,preprintnumbers]{revtex4}

\input{tcilatex}

\begin{document}

\title{Anderson transition in one-dimensional systems with spatial disorder }
\author{Rabah Benhenni $^{1}$ }
\author{ Khaled Senouci $^{1,2}$ }
\email[Corresponding author-Email address: ]{senouci_k@yahoo.com}
\author{ Rachid Bouamrane $^1$ }
\author{ Nouredine Zekri $^{1}$ }
\affiliation{$^{1}$ U.S.T.O., \ D\'{e}partement de Physique L.E.P.M., B.P.1505 El
M'Naouar, Oran, Algeria}
\affiliation{$^{2}$ Universit\'{e} de Mostaganem Abdelhamid Ibn Badis, D\'{e}partement de
Physique, B.P. 227, Route Belhacel, 27000, Mostaganem, Algeria }

\begin{abstract}
A simple Kronig-Penney model for one-dimensional (1D) mesoscopic systems
with $\delta $ peak potentials is used to study numerically the influence of
a spatial disorder on the conductance fluctuations and distribution at
different regimes. We use the L\'{e}vy laws to investigate the statistical
properties of the eigenstates. We found the possibility of an Anderson
transition even in 1D meaning that the disorder can also provide
constructive quantum interferences. We found at this transition that the
conductance probability distribution has a system-size independent shape
with large fluctuations in good agreement with previous works. In these 1D
systems, the metallic phase is well characterized by a Gaussian conductance
distribution. Indeed, the results for the conductance distribution are in
good agreement with the previous works in 2D and 3D systems for other models.
\end{abstract}

\maketitle
\date{\today}

\newpage

\section{Introduction}

\bigskip

The disorder-induced metal-insulator transition (MIT) has been studied
extensively for decades \cite{Erd, Azb} and continues to attract much
attention. Scaling theory \cite{Erd} predicts that all eigenstates of
noninteracting electrons are localized in one-dimensional (1D) and
two-dimensional (2D) systems for any amount of disorder. It is commonly
believed that MIT occurs only for dimensions $d>2$ and the system remain
insulator for $d<2$ \cite{Abr}. However, recently, it was suggested that a
2D Anderson model of localization with purely off-diagonal disorder might
violate this general statement since non-localized states were found at the
band center \cite{Eilm}. It was found that a transition present in this 2D
model can be described by a localization length which diverges with a
power-law behaviour. The possible existence of the Anderson transition in 2D
systems without \ interaction and spin-orbit effects becomes recently a
subject of controversy in the literature \cite{Sus1, Sus2}. \ More recently,
Asada et al. \cite{Asa} studied the $\beta $ function that describes the
scaling of the quantity $\Lambda $ as:

\begin{equation}
\beta (Ln\text{ }\Lambda )=\frac{dLn\text{ }\Lambda }{dLn\text{ }L}
\end{equation}

where $\Lambda $ is the ratio of the quasi-one-dimensional localization
length to the system width for electrons on a long quasi-one-dimensional
system of width $L$. They indicated the possibility of an Anderson
transition for dimensions $d\leq2$ in disordered systems of non-interacting
electrons. On the other hand, \ strong numerical evidence of a mobility edge
was found in disordered photonic systems in two dimensions \cite{Asat}.
Recently, surprising results were found in the properties of disordered
graphene systems \cite{Peres,Fog,Gui,Yan}.The latest studies, both
theoretical and experimental, led to the amazing conclusion that there is no
localization in disordered graphene, even in the one-dimensional (1D)
situation.

It is well known that the conductance $g$ \ is not a self-averaged quantity 
\cite{Was} and therefore fluctuates with the Fermi energy, chemical
potential and sample size. \ Since the conductance does not obey the central
limit theorem \cite{Al1}, it is necessary to investigate not only the first
two moments but the entire probability distribution. Numerical results in 2D
and 3D disordered systems showed that the conductance is normally
distributed in the metallic regime while for strongly localized systems
(insulating regime) a log-normal distribution was found \cite{Shen}. The
exact form of the probability distribution at the transition is not well
known. In such a regime, it was proven that \ the conductance distribution
is independent of the microscopic details of the model (determined by the
distribution of the disorder), of the system size and of the position of the
critical point which separates the metallic and the localized regime in the
phase space of external parameters (energy, disorder). The universality of
the conductance distribution was studied and confirmed by Markos et al. for
2D and 3D models \cite{Mar1, Mar2}. The system-size invariance of the
conductance distribution at the critical points of the MIT was confirmed for
3D and 4D systems in \cite{Mar3,Sle1, Sle2}.

The aim of this paper is to investigate the possibility of an Anderson
transition in 1D spatially disordered systems of noninteracting electron.
Three regimes of electron transport are studied: the insulating regime
corresponding to strong disorder, the metallic regime (corresponding to an
infinitesimal disorder) and the transition regime. In the present work, Levy
statistics \cite{Levy} are used.

\hspace{0.33in}

\section{\ Model description}

We consider a Kronig-Penney model applied to a 1D system of delta potentials
with random positions (spatial disorder) . The corresponding Schr\"{o}dinger
equation can be read:

\begin{equation}
\left\{ -\frac{d^{2}}{dx^{2}}+\sum_{n}\beta \text{ }\delta (x-x_{n})\right\}
\Psi (x)=E\Psi (x)
\end{equation}

\noindent Here $\Psi (x)$ is the single particle wave-function at $x$, $%
\beta $ the potential strength and $E$ \ the single particle energy in units
of $\hbar ^{2}/2m$ with $m$ the electronic effective mass. The two ends of
the system are assumed to be connected ohmically to ideal leads (where the
electron moves freely)

The second order differential equation ($2$) can be mapped by means the
Poincar\'{e} map representation \cite{Sou}:

\begin{equation}
\Psi _{n+1}=\left[ \cos (kl_{n+1})+\frac{\sin (kl_{n+1})}{\sin (kl_{n})}\cos
(kl_{n})+\beta _{n}\frac{\sin (kl_{n+1})}{k}\right] \Psi _{n}-\frac{\sin
(kl_{n+1})}{\sin (kl_{n})}\Psi _{n-1}
\end{equation}

\noindent where $\Psi _{n}$ is the value of the wave-function at site $n$
and $k=\sqrt{E}$ is the electron wave number $\ l_{n+1}=x_{n+1}-x_{n}$ is
the inter-atomic spacing. $l_{n+1}=a+s$, $s$ being a random variable
uniformly distributed as $-W/2<s<W/2$ ($W$ being the degree of the disorder).

For equally spaced potentials, the interatomic spacing $l_{n+1}=l_{n}=a$ ($%
W=0$) and equation $(2)$ reduces to that found by Soukoulis et al.\cite{Sou}:

\begin{equation}
\Psi _{n+1}+\Psi _{n-1}=\left[ 2\cos \sqrt{E}+\beta _{n}\frac{\sin \sqrt{E}}{%
\sqrt{E}}\right] \Psi _{n}
\end{equation}

The solution of equation ($3$) is carried out iteratively by taking the two
initial wave functions at sites $1$ and $2$ : $\Psi _{1}=$ $\exp (-ik)$ and $%
\Psi _{2}=$ $\exp (-2ik)$. We consider here an electron having a wave number 
$k$ incident at site $N+3$ from the right (by taking the chain length $L=N$,
i.e. $N+1$ scatterers). The transmission coefficient ($T$) reads

\begin{equation}
T=\frac{|1-exp(-2ikl_{n})|^{2}}{|\Psi _{N+2}-\Psi _{N+3}exp(-ikl_{n})|^{2}}
\end{equation}

The dimensionless conductance ($g=\frac{G}{e^{2}/h}$ ) can be obtained from
the transmission coefficient $T$ via the Landauer formula for 1D systems 
\cite{Lan}:

\begin{equation}
g=\frac{2T}{1-T}
\end{equation}%
where the factor two arises from the two possible states of the electron
spin.

and the variance of conductance ($Ln(g)$) reads:

\begin{equation}
var(Ln(g))=\text{ }<Ln(g)^{2}>-<Ln(g)>^{2}
\end{equation}

where $<...>$ denotes an average over different realizations of the disorder.

\section{Results and discussion}

In this section, we discuss numerical results of the conductance probability
distribution for different amount of spatial disorder of 1D mesoscopic
systems of delta potentials. In order to obtain the probability distribution
of the conductance, we build a statistical ensemble of $10^{4\text{ }}$%
samples differing only in the realization of the disorder.

Firstly, we examine the effect of this kind of disorder on electronic
eigenstates by calculating the transmission coefficient and the conductance
probability distribution. Figure 1 shows the scaling of the transmission
coefficient $<-Ln(T)>$ for different degrees of spatial disorder and for two
values of electron energy $E=4$ (corresponding to an energy in the band gap
of the periodic system) and $E=12$ (corresponding to an energy in the
allowed band, see inset of \ Figure 1.a). For strong disorder the
transmission coefficient decreases exponentially with the length scale.
Furthermore, the localization length $\zeta $ and the Lyapunov exponent $%
\gamma =1/\zeta $(slope of the curve) are deduced. The calculated results
are presented in table I. Fig.1a shows that increasing disorder leads to
strong localization (since the localization length $\zeta $\ decreases with
the disorder strength $W$). This is in good agreement with previous results
with topological disorder i.e the disorder localizes the electronic
eigenstates. However, this is not the case as shown in Fig.1b where $\zeta $
increases with the disorder. This behaviour was previously pointed out by
Nimour et al. for 1D spatially disordered systems with finite width
potentials \cite{Nim}. More recently, it has been shown that the disorder
could either suppress or enhance the transmission in disordered grapheme
superlattices \cite{Blio}.

\ \bigskip

\begin{table}[t]
\begin{center}
\begin{tabular}{||l||l||l||l||l||}
\hline\hline
& $E=4$ & $E=4$ & $E=12$ & $E=12$ \\ \hline\hline
$W$ & $\zeta $ & $\ \gamma $ & $\zeta $ & $\ \gamma $ \\ \hline\hline
$0.005$ & $4797.866$ & $2.084\times 10^{-4}$ & $1.7832$ & $0.5607$ \\ 
\hline\hline
$0.1$ & $1205.313$ & $8.2966\times 10^{-4}$ & $1.8082$ & $0.5530$ \\ 
\hline\hline
$0.2$ & $0301.204$ & $0.00332$ & $1.9205$ & $0.52068$ \\ \hline\hline
\end{tabular}%
\end{center}
\caption{Calculated results of the Lyapunov exponent $\protect\gamma $ and
the localization length $\protect\zeta $ as a function of the disorder$\ W.$}
\end{table}

Since our system is disordered, its conductance depends on specific
realizations of disorder (for a given $L$), it is then appropriate to study
the whole probability distribution of conductance. For strong disorder where 
$w=0.2$, we have $\zeta =300.2$, which is smaller than the size of the
system. The localization of the electronic eigenstates is confirmed in
figure 2a where the probability distribution of the natural logarithm of the
conductance $(Ln(g))$ is plotted. \ In this figure the probality
distribution is Gaussian indicating the general behaviour in this regime 
\cite{Shen}. This figure shows also that the mean conductance decreases as
the system size increases. In a previous interesting paper, Vagner et al. 
\cite{Vag} studied analytically electron transport in a one-dimensional wire
with disorder modeled as a chain of randomly positioned scatterers with $%
\delta $-shaped impurity potential (similar to our model). They found that
the distribution $P(f)$ of the variable $f=ln(1+\rho )$ ($\rho $ being the
resistance) has a non-Gaussian behaviour in the limit of weak disorder. This
result is confirmed for the conductance in our system. It is clearly seen
from figure 2b that when the disorder decreases, the conductance
distribution shows a deviation from its log-normal form indicating
delocalization of the eigenstates.

In order to further understand the behaviour of the conductance, we
investigate the conductance distribution $P(g)$ for different amounts of
spatial disorder and for different system sizes. All these distributions
show long power law tails decreasing for large values of the conductance $g$
(see Figure 3). Therefore, we use the Levy statistics $L_{\mu }(Z)$ of index 
$\mu $ which decreases as $Z^{-(1+\mu )}$ for large values of Z \cite{Levy}.
It is found that the conductance distribution $P(g)$ behaves as $g^{-(1+\mu
)}$ for large values of $g$. The exponent $\mu $ is then extracted from the
log-log plot of $P(g)$ for large values of $g$ which is linear with a slope
equal to $-(1+\mu )$ (see the inset). It is known that if \ $\mu >2$, the
probability distribution is normal. On the other hand, if $\mu <2$ \ it
means that the distribution is log-normal. In figure 4, the index $\mu $ is
plotted as a function of degree of spatial disorder $W$ and for different
system size $L$. In this figure two distinguishable regions are shwn: a
region of $\mu <2$ (corresponding to an insulating regime) and a region of $%
\mu >2$ (corresponding to a metallic regime for small disorder $W$). The
intersection point of curve $\mu (W)$ with the strait line $\mu =2$
corresponds to the metal-insulator transition region. The critical point
separating the metallic and insulating regimes corresponds to a critical
disorder with\ $W_{c}=0.0035$. Figure 4 shows also that the index $\mu $\ is
independent of the system size at the transition.

Let us now examine the conductance probability distribution for different
amounts of spatial disorder. In Figure 5 are plotted the probability
distributions of $Ln(g)$ (Fig.5a) and $g$ (Fig.5b) for different system size
($L=500,700,800$ and $900$) for the critical disorder $W_{c}=0.0035$. The
conductance distribution seems to be neither normal nor log-normal. The size
independence of the conductance distribution is in agreement with the
observed one for 2D and 3D systems \cite{Mar1, Mar3, Sle2} at the
transition. The long tail of the distribution in Fig.5a is representative
for large fluctuations in good agreement with the results of Shapiro \cite%
{Sha1} and Shapiro and Cohen \cite{Sha2} for the metal-insulator transition.

For an infinitesimal disorder $W<W_{c}$ corresponding to $\mu >2,$ where $%
w=0.0025$, we have $\zeta =47234$, which is much larger than the system size 
$L$. The conductance distribution has a Gaussian form indicating the
metallic regime (Figure 6).\bigskip\ This figure shows the general behaviour
for this regime. Once the system is away from the critical point, the
conductance distribution $P(g)$ begins to show size dependence. Distribution 
$P(g)$ moves toward higher values of the conductance as the system size
increases.

\section{Conclusion}

We have used the Kronig-Penney model in a simple one-dimensional (1D) system
with spatial disorder to determine the metal-insulator transition and
examine the size independence of its distribution at this transition. To
find the critical point, we have used stable Levy laws \cite{Levy}. We found
the possibility of an Anderson transition even in 1D meaning that the
disorder can also provide constructive quantum interferences. We found the
critical disorder $W_{c}$ for this transition. Indeed, the results are in
good agreement with the previous works in 2D and 3D systems for other models 
\cite{Mar1, Mar3, Sle2} for metal-insulator transition. The conductance
probability distribution is found to be Gaussian in the metallic regime. At
the transition, the distribution $P(lng)$ exhibits the typical asymmetric
behaviour while in the localized regime, the distribution is log normal. It
is important to study the universality of the conductance distribution at
the transition in a system of finite width potentials where the conductance
fluctuations are less important in comparison to the present model \cite{Cot}%
, and to find the critical points of the transition in the energy-disorder
phase space and for different kinds of disorder. It is also important to
investigate the effect of spatial disorder on conductance fluctuations in
different regimes. These problems will be the subject of a forthcoming paper.

\section*{Acknowledgments}

One of us (K.S.) gratefully acknowledges the partial support from Physics
Department of the University of Mostaganem (Algeria). K.S. would also like
to thank the International Centre for Theoretical Physics (ICTP), Trieste
(Italy) for its hospitality during the period when part of this work was
carried out. This work has been supported by Algerian National Research
Project CNEPRU (D02220080025 approved in 2008).

\textbf{\newpage }

\bigskip\ \ \ \ \ \ \ \ \ \ \ \ \ \ \ \ \ \ \ \ \ \ \ \ \ \ \ \ \ \ \ \ \ \
\ \ \ \ \ \ \ \ \ \ \ \ \ \ \textbf{Figure captions}

\textbf{Figure 1: }Transmission coefficient \TEXTsymbol{<}-Ln(T)\TEXTsymbol{>%
} as a function of system size averaged over $1000$ realizations of the same
system for $V=2$ and different amounts of spatial disorder $W$ and for a) $%
E=4$ (in the miniband). b) $E=12$ (in the gap). Inset in a: Transmission Vs
energy for a periodic system.

\textbf{Figure 2:} Probability distribution of $-log(g)$ for $E=4$, $V=2$, $%
L=1600$ and for a) $W=0,2$. b) $W=0,0036$.

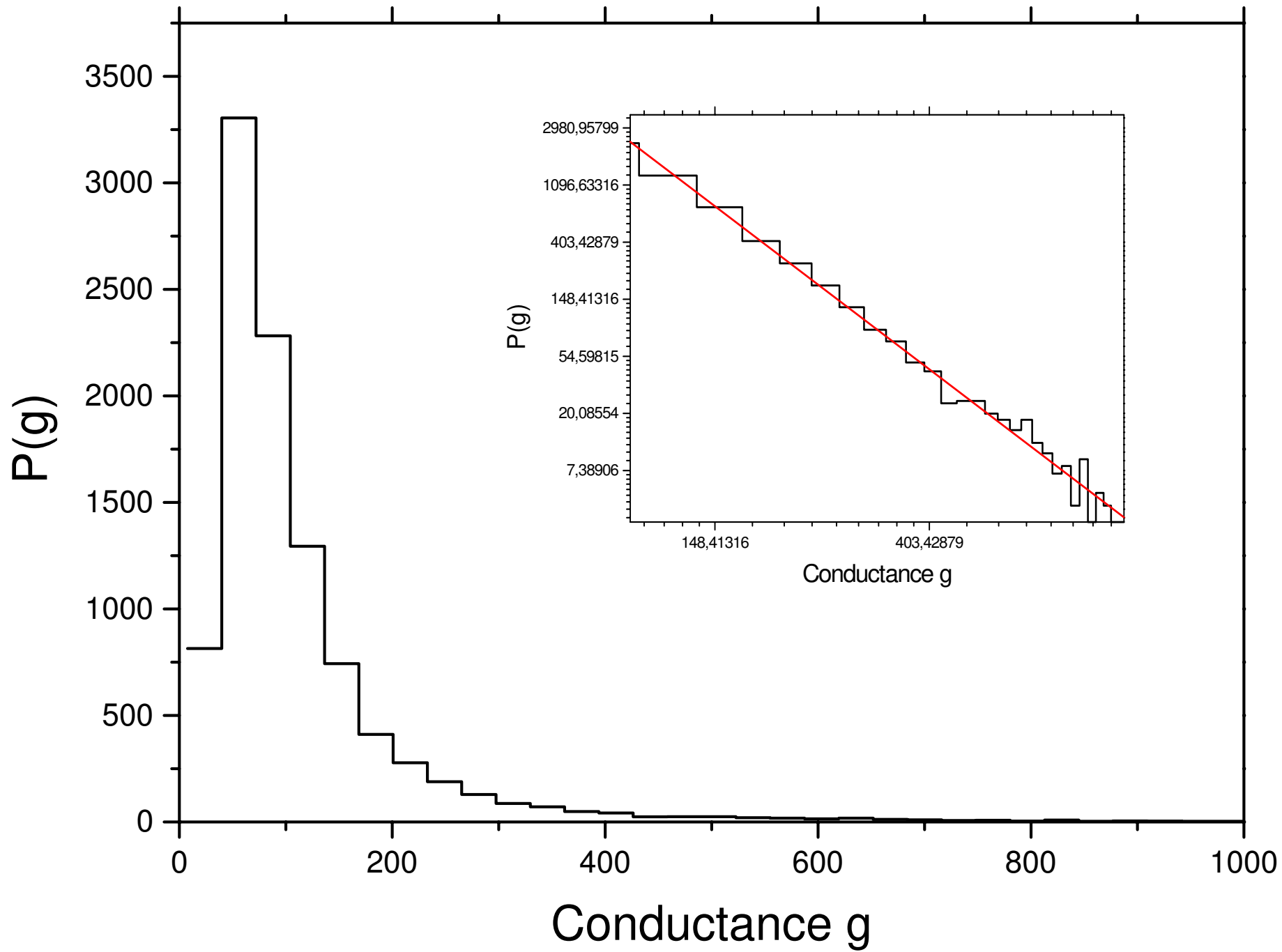
\textbf{Figure 3: }Conductance probability distribution for $L=1500$, $E=4$
and $W=0.0034$. Inset: log-log plot of the tail of the distribution, the
strait line corresponds to the best power-law fit to the tail $P(g)\sim
g^{-(1+%
\mu
)}$ with $%
\mu
=1.02.$

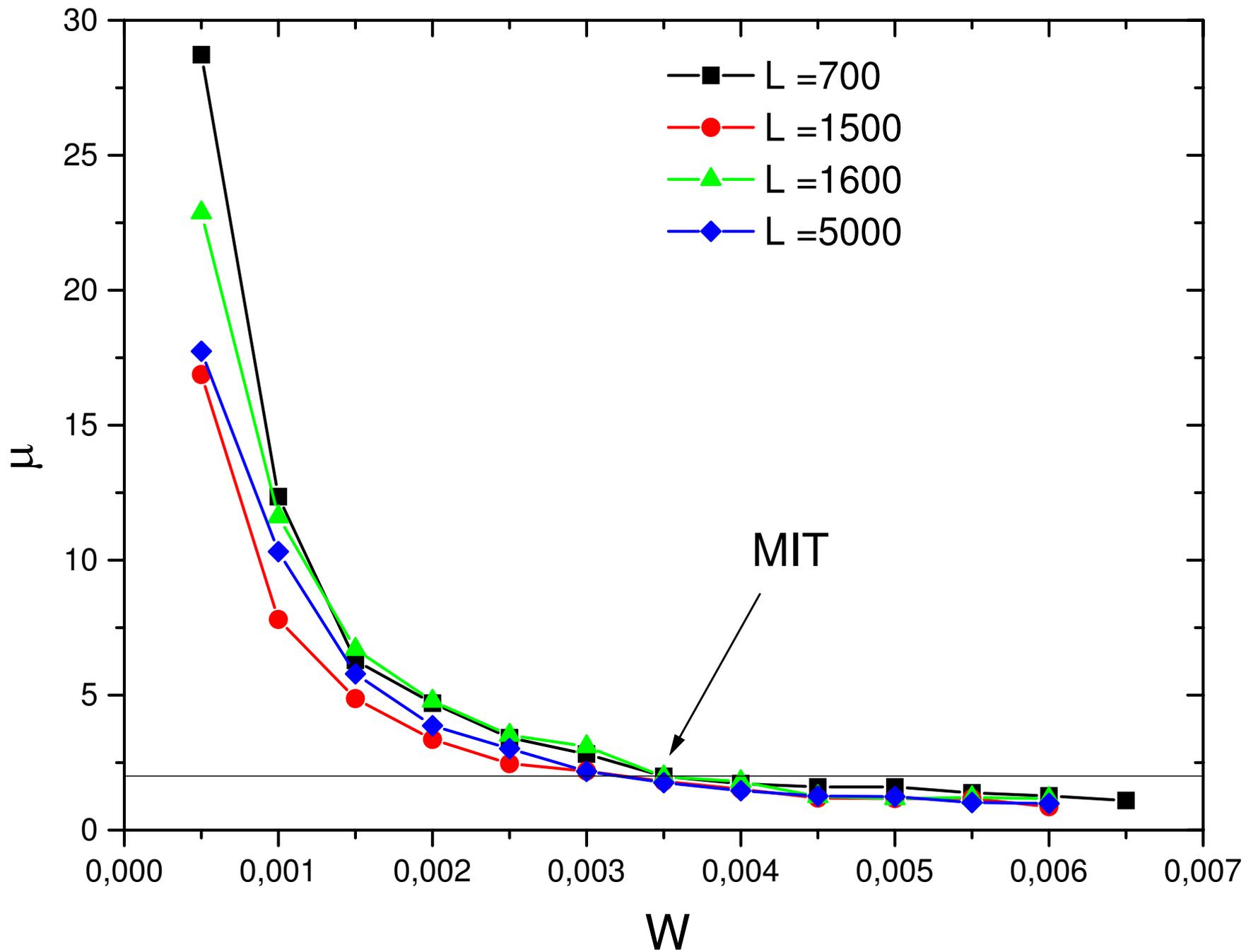
\textbf{Figure 4: }L\'{e}vy exponent $\mu $\ as a function of disorder for $%
E=4$, $V=2$ and different system size $L$.

\textbf{Figure 5: }a) Probability distribution of the conductance $g$ for $%
E=4$ and $V=2$. b) Probability distribution of $-\ln (g)$ compared with a
Gaussian with the same mean and variance.

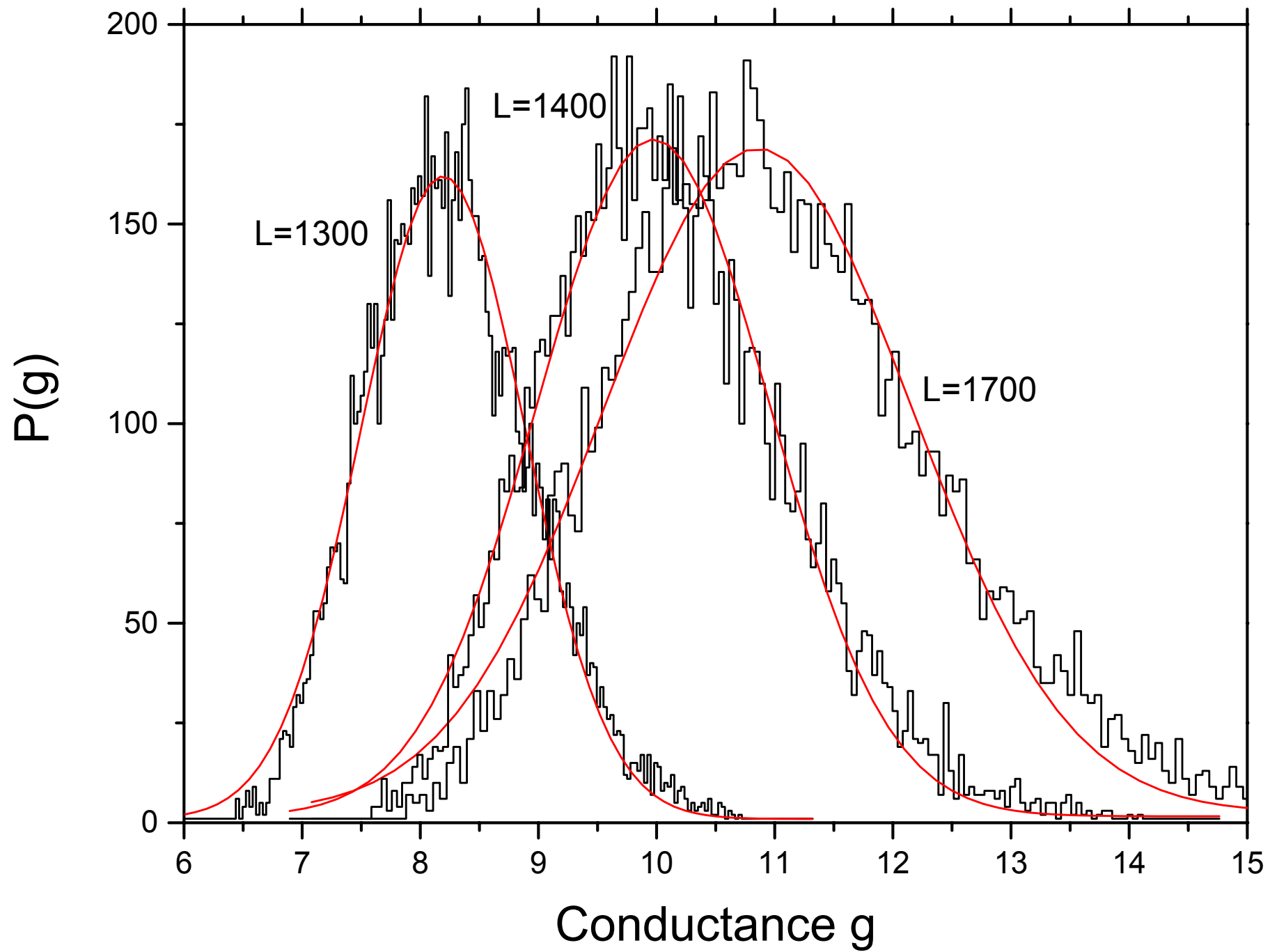
\textbf{Figure 6: }Conductance probability distribution in the metallic
regime for different system size $L$ with $W=0.0025$, $E=4$ and $V=2$
compared with a Gaussian with the same mean and variance.

\bigskip


\newpage

\bigskip

\begin{thebibliography}{99}
\bibitem{And} P. W. Anderson, Phys. Rev. \textbf{109} \ (1958) 1492 .

\bibitem{Erd} P. Erd\"{o}s and R. C. Herndon, Adv. Phys\textit{.} \textbf{31}
\ (1982) 65.

\bibitem{Azb} M. Y. Azbel and P. Soven, Phys. Rev. B \textbf{27} (1983) 831 .

\bibitem{Abr} E. Abrahams, P.W. Anderson, D.C. Licciardello and T.V.
Ramakrishnan, Phys. Rev. Lett. \textbf{42} (1979)\ 673.

\bibitem{Eilm} A. Eilmes , R. A. Romer \ and M. Schreiber, Physica B \textbf{%
296} (2001)\ 46.

\bibitem{Sus1} I. M. Suslov, Zh. Eksp. Teor. Fiz. \textbf{128} (2005) 768
[JETP \textbf{101} (2005) 661 ]; cond-mat/0504557.

\bibitem{Sus2} I. M. Suslov, Zh. Eksp. Teor. Fiz. \textbf{129} (2006) 1064
[JETP \textbf{102} (2006) 938 ]; cond-mat/0512708.

\bibitem{Asa} Y. Asada, K. Slevin, and T. Ohtsuki, Phys. Rev. B \textbf{73}
(2006) 041102 (R).

\bibitem{Asat} A. A. Asatryan, L. C. Botten and M. A. Byrne, Phys. Rev. E 
\textbf{75} (2007) 015601 (R).

\bibitem{Peres} N. M. R. Peres, F. Guinea, and A. H. Castro Neto, Phys. Rev.
B \textbf{73} (2006) 125411.

\bibitem{Fog} M. M. Fogler, D. S. Novikov, I. I. Glazman, and B. I.
Shklovskii, Phys. Rev. B \textbf{77} (2008) 075420 .

\bibitem{Gui} F. Guinea, M. I. Katsnelson, and M. A. H. Vozmediano,
Phys.Rev. B \textbf{77} (2008) 075422.

\bibitem{Yan} X.-Z. Yan and C. S. Ting, Phys. Rev. Lett. \textbf{101} (2008)
126801.

\bibitem{Was} S.Washburn and R.A.Webb, Adv. Phys. \textbf{35} (1986) 375.

\bibitem{Al1} B .L. Alt'shuler, A.G.Aronov and B. Z. Spivak, Sov.Phys. JETP
Lett. \textbf{33} (1981) 94.

\bibitem{Shen} P. Sheng and Z. Zhang, J. Phys. Condens. Matter \textbf{3}
(1991) 4257 .

\bibitem{Mar1} P. Markos, Europhys. Lett. \textbf{26} 431 (1994); J. Phys.I
France \textbf{4} (1994) 551.

\bibitem{Mar2} P. Markos and B. Kramer, Phil. Mag. B \textbf{68} (1993) 357.

\bibitem{Mar3} P. Markos, Phys. Rev. Lett. \textbf{83} (1999)588 ; M. R\"{u}%
hl\"{a}nder, P. Markos and C. M. Soukoulis, Phys. Rev. B \textbf{64} (2001)
212202.

\bibitem{Sle1} K. Slevin and T. Ohtsuki, Phys. Rev. Lett.\textbf{78} (1997)
4083.

\bibitem{Sle2} K. Slevin and T. Ohtsuki, Phys. Rev. Lett. \textbf{82} (1999)
669.

\bibitem{Levy} P. L\'{e}vy, \textit{Th\'{e}orie de l'Addition des Variables
Al\'{e}atoires }(Gautier-Villars, Paris, 1937); M.Shlesinger, G. M.
Zaslavsky and U. Fish, \textit{L\'{e}vy Flights and Related Topics in Physics%
} (Springer, Berlin, 1995); J.P. Bouchaud and A. Georges, Phys. Rep. \textbf{%
195} (1990) 12.

\bibitem{Sou} C. M.Soukoulis, J.V. Jos\'{e}, E. N. Economou and P. Sheng,
Phys. Rev. Lett. \textbf{50} (1983) 764.

\bibitem{Lan} R. Landauer, Phil. Mag. \textbf{21} (1970) 263.

\bibitem{Nim} S. M. A. Nimour, R. Ouasti and N. Zekri, Phys. Stat.Sol. (b) 
\textbf{209} (1998) 311.

\bibitem{Blio} Y. P. Bliokh,V. Freilikher, S. Savel'ev and F. Nori, Phys.
Rev. B \textbf{79} (2009) 075123.

\bibitem{Vag} P. Vagner, P. Markos , M. Mosko and T. Schapers, Phys. Rev. B 
\textbf{67} (2003) 165316 .

\bibitem{Sha1} B. Shapiro, Phil. Mag. B \textbf{56 }(1987) 1032; Phys. Rev.
Lett. \textbf{65} (1992) 595.

\bibitem{Sha2} B. Shapiro and A.Cohen, Int. J. Mod. Phys. B \textbf{6}
(1992) 1243.

\bibitem{Cot} E. Cota, J.V. Jose and M.Ya. Azbel, Phys. Rev. B \textbf{32}
(1985) 6157.
\end{thebibliography}
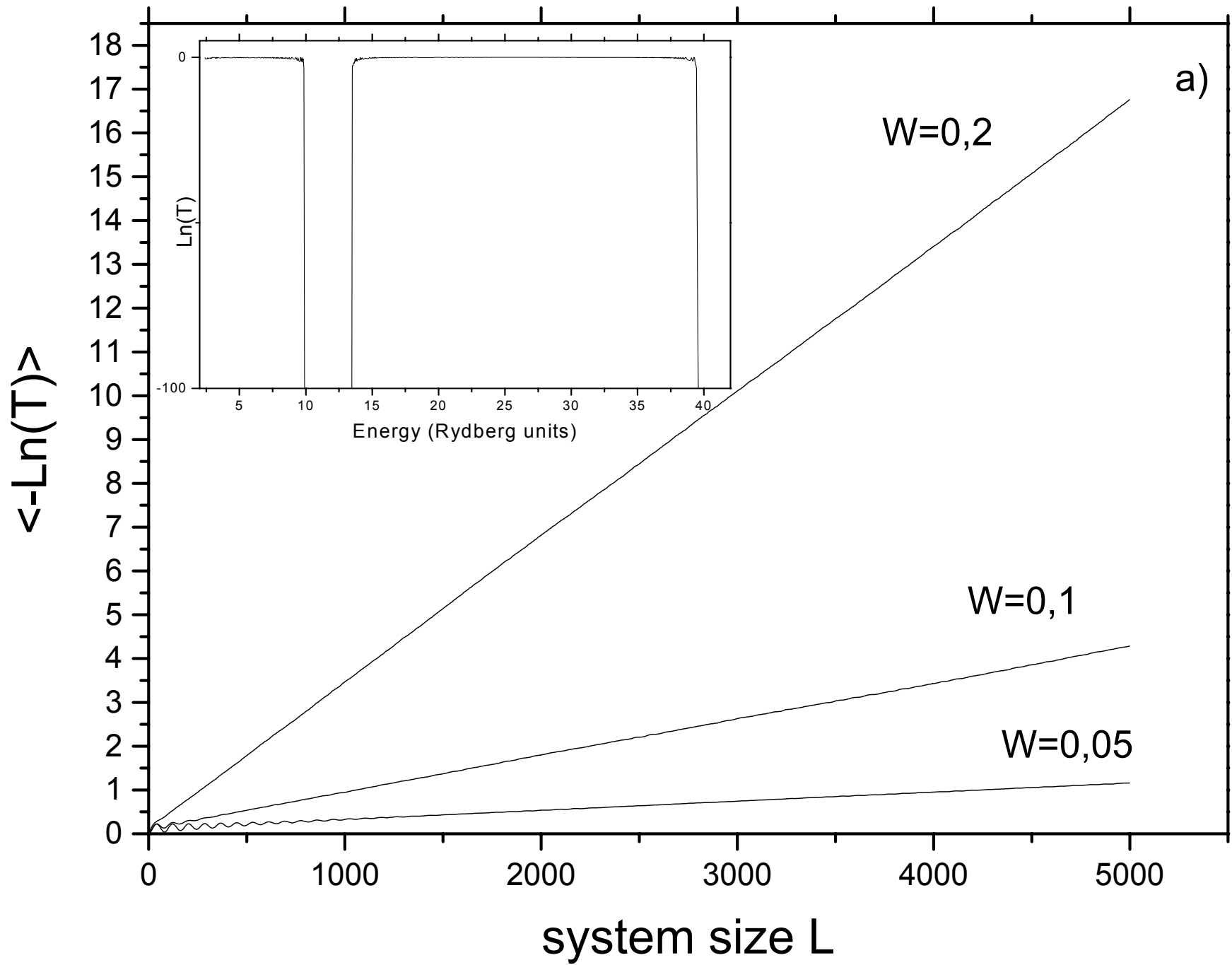
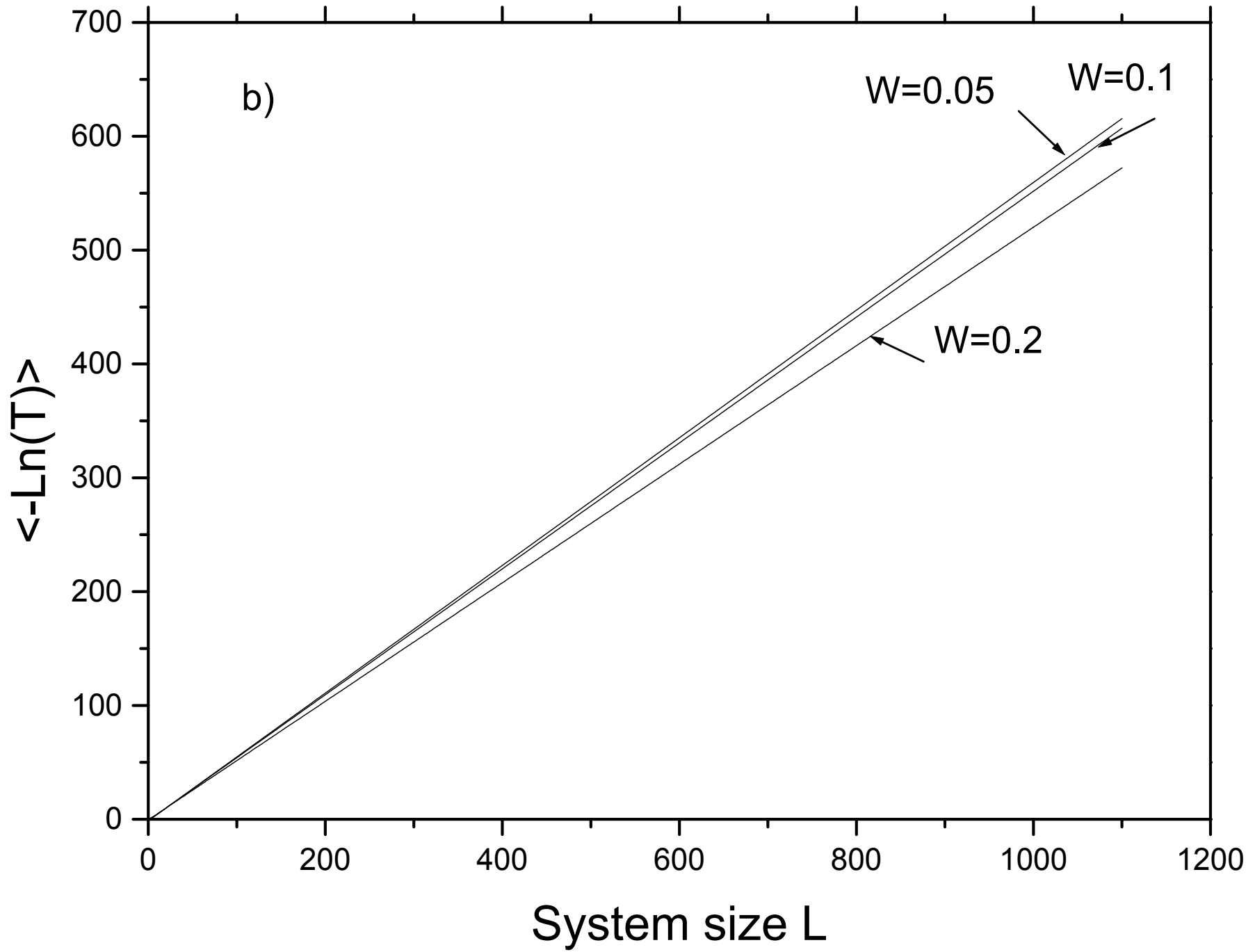
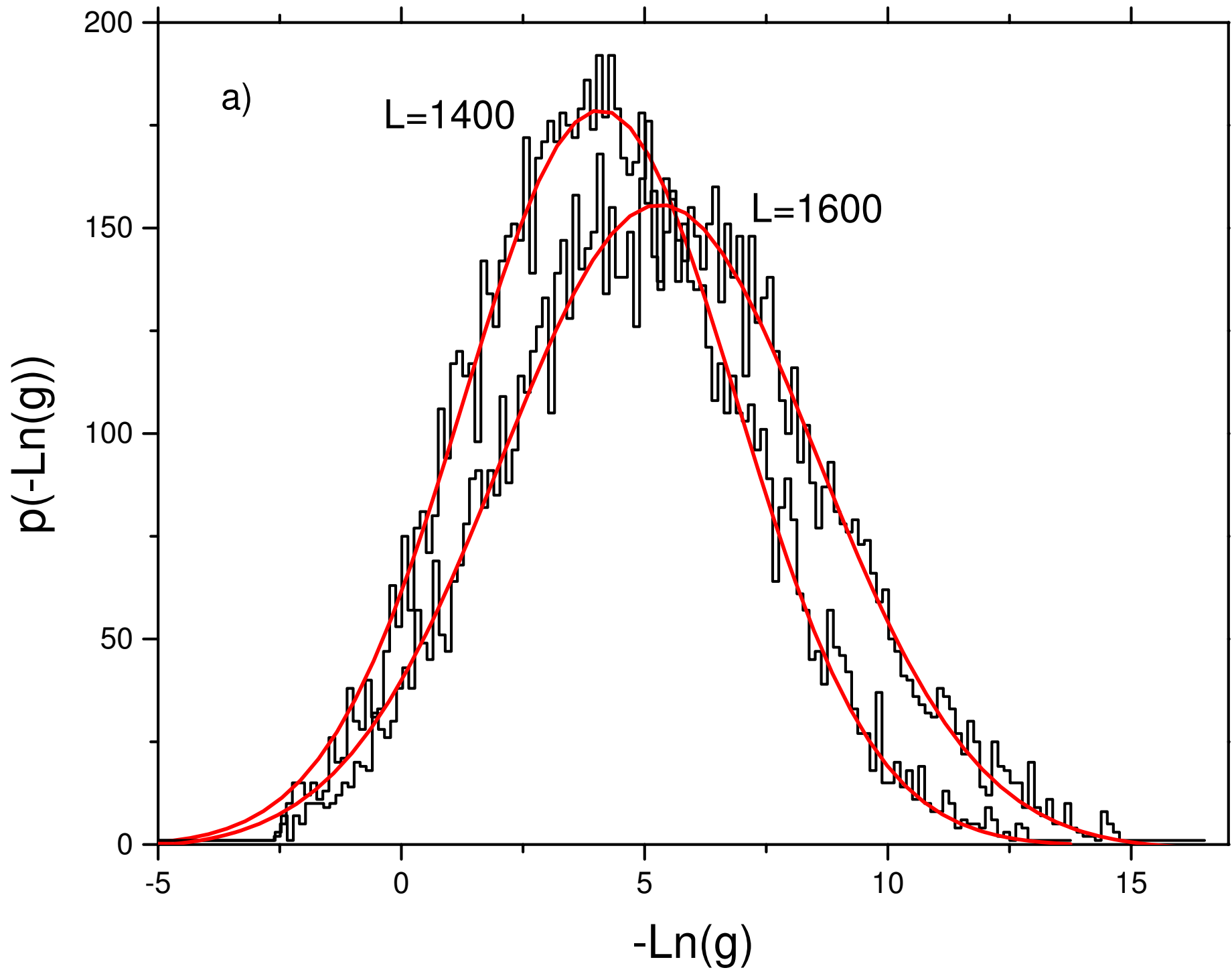
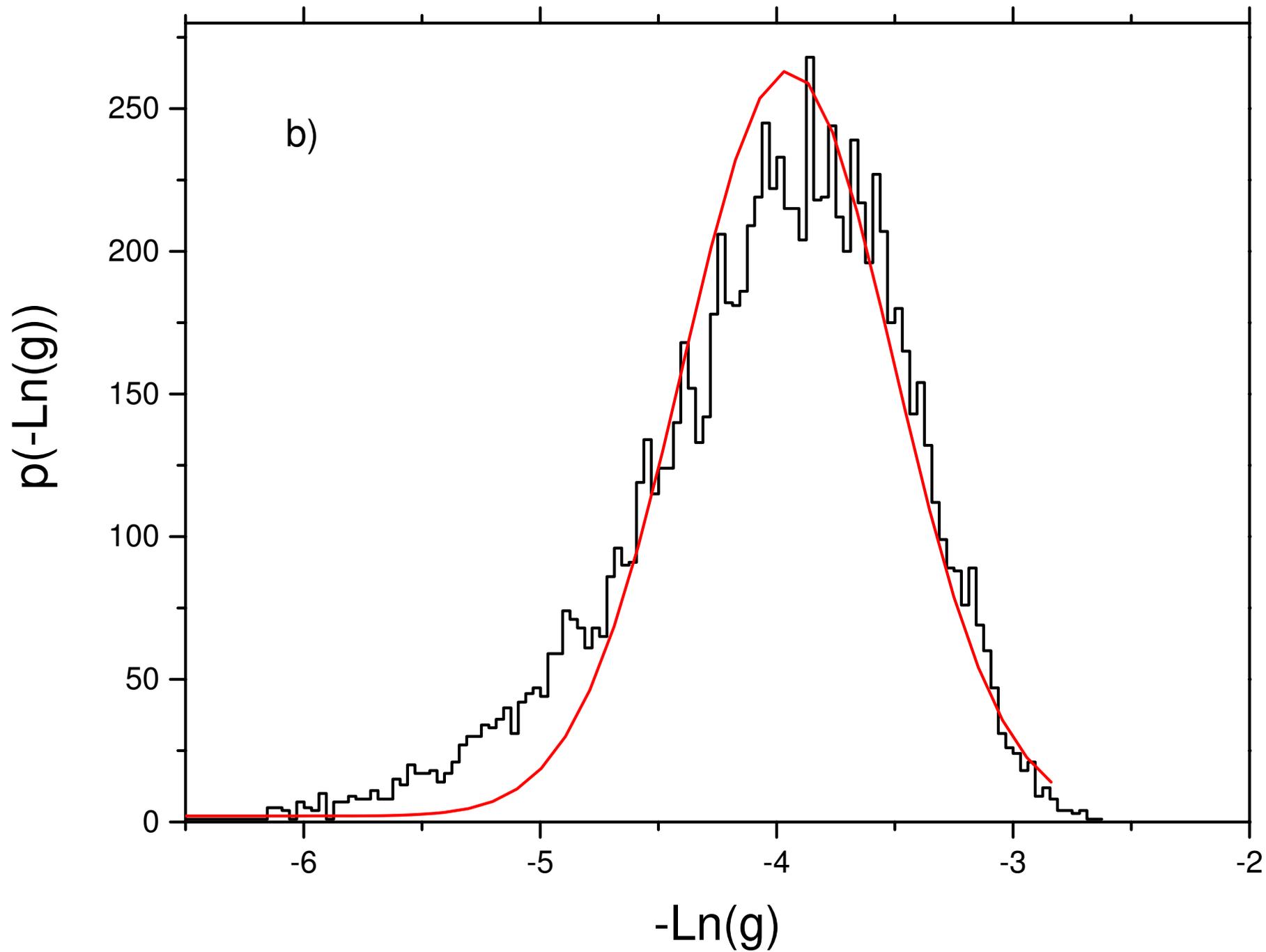
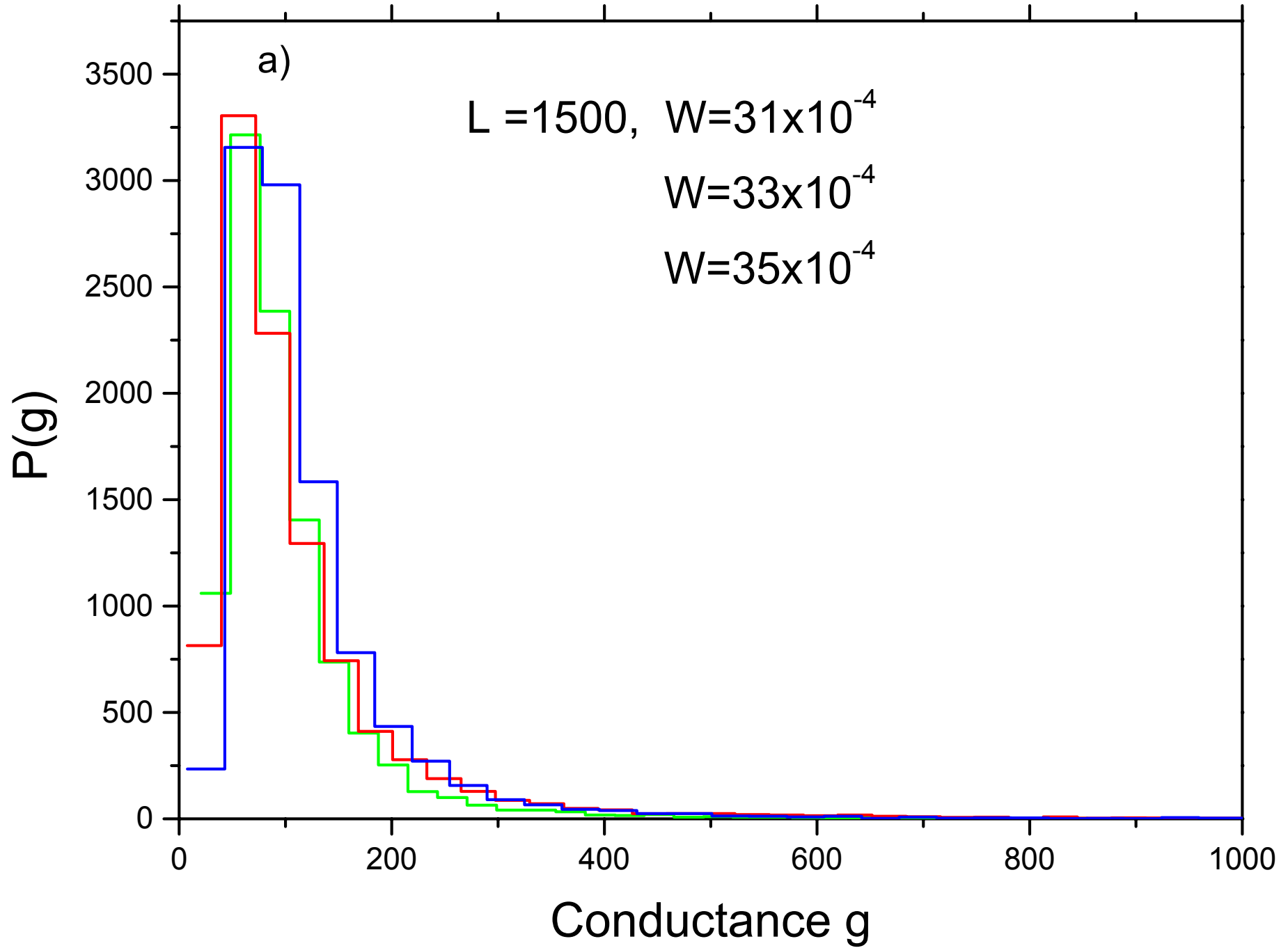
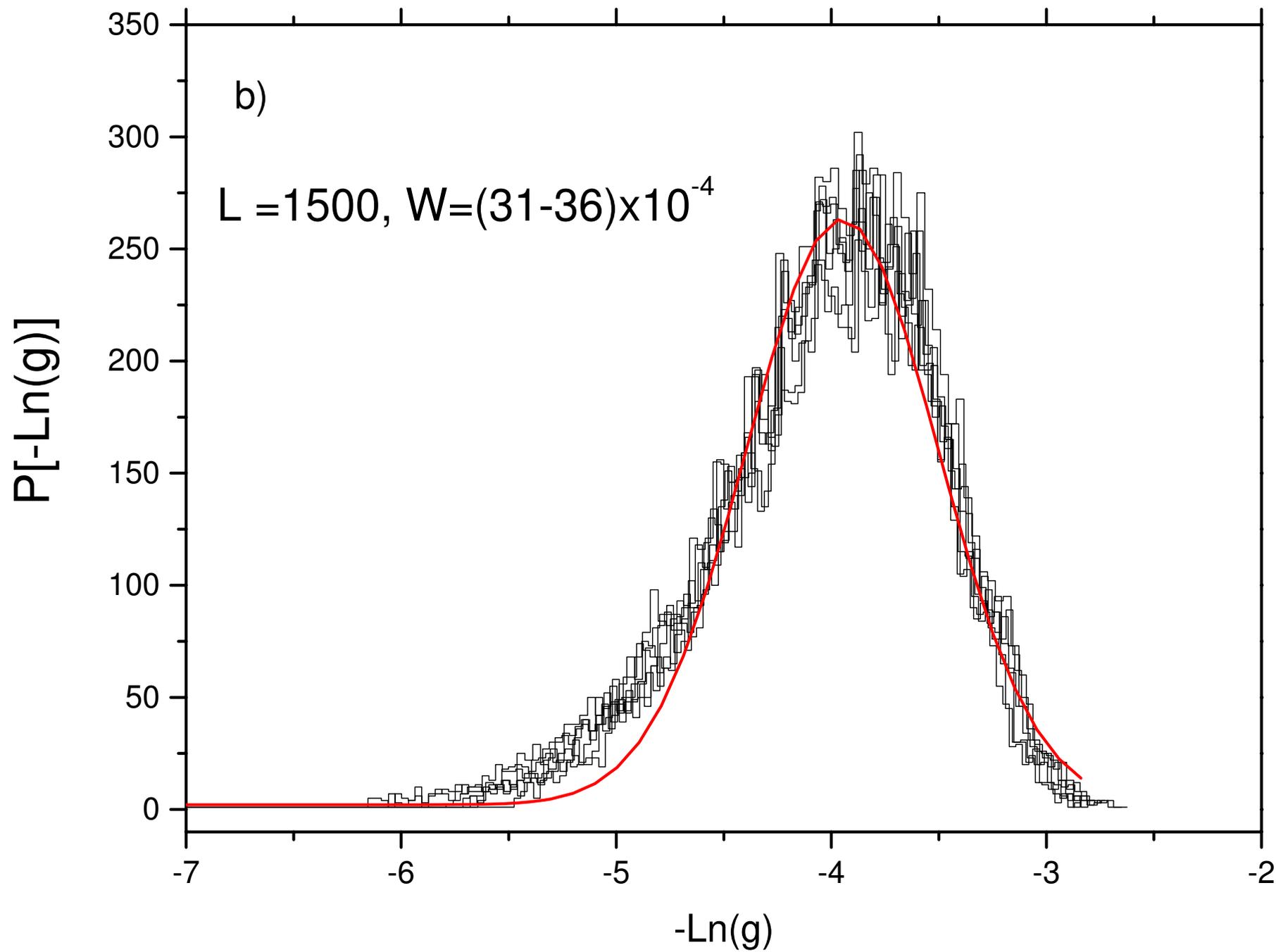
\end{document}